\documentclass[aps,preprint,showpacs,superscriptaddress,groupedaddress]{revtex4}  
\usepackage{graphicx}  
\usepackage{subfigure} 
\usepackage{dcolumn}   
\usepackage{bm}        
\usepackage{amssymb}   
\usepackage{slashed}   
\usepackage{amsmath}   
\usepackage{simplewick} 
\usepackage{verbatim}  
\usepackage{color}

\begin{document}

\widetext


\title{\large \bfseries \boldmath Effect of the Charged Higgs Bosons in the Radiative Leptonic Decays of $B^-$ and $D^-$ Mesons}
\author{Ji-Chong Yang}
\author{Mao-Zhi Yang} 
\affiliation{School of Physics, Nankai University, Tianjin 300071, P.R. China}

\vskip 0.25cm

\date{\today}

\begin{abstract}
In this work, we study the radiative leptonic decays of $B^-$ and $D^-$ mesons in the standard model and the two-Higgs-doublet-model type II. The results are obtained using the factorization procedure, and the contribution of the order $O(\Lambda _{\rm QCD}\left/m_Q\right.)$ is included. The numerical results are calculated using the wave-function obtained in relativistic potential model. As a result, the decay mode $B\to \gamma \tau \nu_{\tau}$ is found to be sensitive to the effect of the charged Higgs boson. Using the constraint on the free parameters of the two Higgs doublet model given in previous works, we find the contribution of the charged Higgs boson in the decay mode $B\to \gamma \tau \nu_{\tau}$ can be as large as about $13\%$.
\end{abstract}

\maketitle

\section{\label{sec:level1}Introduction}

A resonance at $126\;{\rm GeV}$ is discovered in both the ATLAS~\cite{higgs1} and CMS experiments~\cite{higgs2}, which is consistent with the Standard Model (SM) Higgs boson~\cite{higgs3}. This is one of the great experimental achievement in recent years. Despite the great success of SM, there are still problems implying new physics beyond SM. One of the simplest extensions is the two Higgs doublet model (2HDM), with a second Higgs doublet introduced by Ref.~\cite{2hdm_start} (see also \cite{2hdm_review}). The Higgs potential and the Yukawa Lagrangian corresponding to the 2HDM is not unique. The one with a CP-invariant potential, and with the Yukawa Lagrangian such that one of the Higgs doublet couples to the down sector of fermions while the other couples to the up, which is so called 2HDM-Type-II~\cite{typeii}. Such a Lagrangian can avoid tree-level flavor changing neutral currents (FCNCs)~\cite{fcnc}. There are 2 Vacuum Expectation Values (VEV), denoted as $v_1$, and $v_2$ in 2HDM, and 5 Higgs bosons denoted as $h$, $A$, $H^0$ and $H^{\pm}$. By studying the phenomenons induced by 2HDM, and comparing the results with the experimental data, one can detect the signal of the existence of new physics, and fix the free parameters in the model.

The radiative leptonic decay of the heavy pseudoscalar meson with a massive lepton provides a good opportunity to study the 2HDM. This process can be mediated both by $W$ bosons and $H^{\pm}$ bosons, however, the contribution of the charged Higgs bosons is barely investigated before. The contribution of 2HDM is usually small. To probe the signal, sufficient precision of the numerical results in the SM is required. In this decay mode, the strong interaction is involved only in the initial hadronic state. As a result, this decay mode has been investigated using the factorization approach~\cite{Sachrajda,factorization}, and results has been calculated up to the order $O(\left.\alpha _s\Lambda _{\rm QCD}/m_Q\right)$ in the SM~\cite{factorization}. Apart from that, with a massive lepton in the final state, the results are expected to be sensitive to the charged Higgs bosons.

There are two free parameters in 2HDM associated with this decay mode, $\tan \beta \equiv \left. v_2/\right. v_1$, and $M_{H^{\pm}}$. The branching ratios can be expressed as the function of $R \equiv \left. \tan \beta / M_{H^{\pm}}\right.$. Using the constraints on those parameters~\cite{RConstraint1,mhConstraint1,mhConstraint2,tanbetaConstraint}, the branching ratios including the contribution of the charged Higgs bosons can be obtained. The results are shown in Table~\ref{brtot}. We find that, the decay mode $B\to \gamma \tau \nu_ {\tau}$ is very sensitive to the charged Higgs bosons, which makes it a good decay mode to probe the signal of 2HDM. On the other hand, the decay mode $D\to \gamma \mu \nu_ {\mu}$ is barely affected by 2HDM, and should be excluded in the search of the charged Higgs bosons.

This paper is organized as follows. Sec.~II is a briefly review of the factorization in the SM. The contribution of the 2HDM is discussed in Sec.~III. The numerical results and analysis are contained in Sec.~IV. Sec.~V is a summary.

\section{\label{sec:level2}Factorization in Standard Model}

\begin{figure}
\includegraphics[scale=0.8]{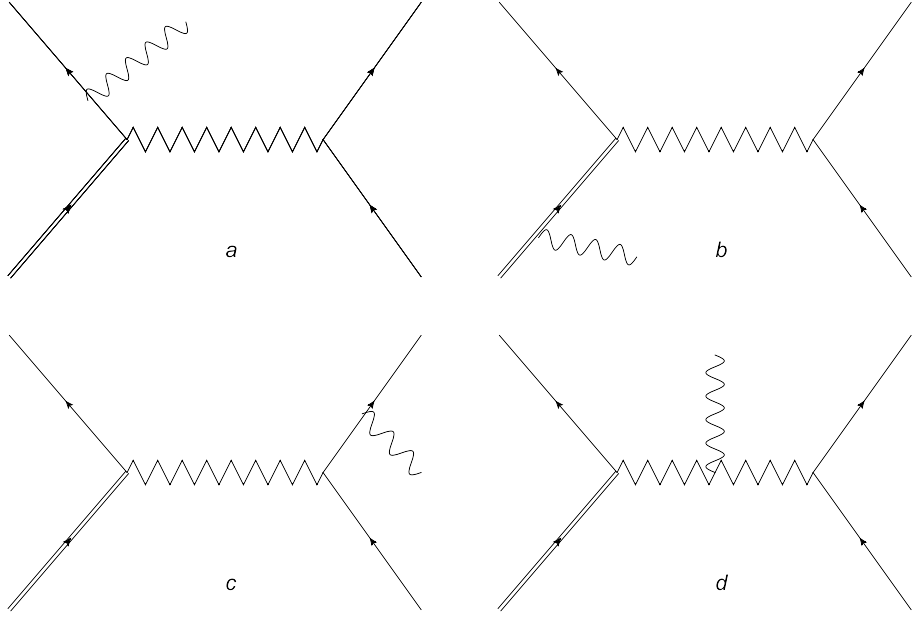}
\caption{\label{Tree1} tree level amplitudes in SM, the double line represents the heavy quark propagator but not the HQET propagator.}
\end{figure}

The heavy pseudoscalar meson is constituted with a quark and an anti-quark, and one of the quarks is a heavy quark. The Feynman diagrams of the radiative leptonic decay of $B$ meson at tree level can be shown as Fig.~\ref{Tree1}. The contribution of Fig.~\ref{Tree1}.~d is suppressed by a factor of $1\left/M_w{}^2\right.$, and can be neglected. The amplitudes of Fig.~\ref{Tree1}.~a, b and c can be written as
\begin{equation}
\begin{split}
&\mathcal{A}_a^{(0)}=\frac{- i e_q G_F V_{Qq}}{\sqrt{2}}\bar{q}(p_{\bar{q}})\slashed \varepsilon _{\gamma }^*\frac{\slashed p_{\gamma }-\slashed p_q}{2p_{\gamma}\cdot p_{\bar{q}}}P_L^{\mu } Q(p_Q)\left(\bar{l}P_{L\mu}\nu\right)\\
&\mathcal{A}_b^{(0)}=\frac{- i e_Q G_F V_{Qq}}{\sqrt{2}}\bar{q}(p_{\bar{q}})P_L^{\mu }\frac{\slashed p_Q-\slashed p_{\gamma}+m_Q}{2p_Q\cdot p_{\gamma}}\slashed \varepsilon _{\gamma }^* Q(p_Q)\left(\bar{l}P_{L\mu}\nu\right)\\
&\mathcal{A}_c^{(0)}=\frac{ -e G_F V_{Qq}}{\sqrt{2}}\bar{q}(p_{\bar{q}})P_L^{\mu } Q(p_Q)\left(\bar{l}\slashed \varepsilon _{\gamma }^*\frac{i(\slashed p_{\gamma }+\slashed p_l+m_l)}{2\left(p_{\gamma }{\cdot}p_l\right)}P_{L\mu}\nu\right)
\end{split}
\label{eq.1.1}
\end{equation}
where $p_{\bar{q}}$ and $p_Q$ are the momenta of the anti-quark $\bar{q}$ and quark $Q$, $p_{\gamma}$, $p_l$ and $p_{\nu}$ are the momenta of photon, lepton and neutrino, $\varepsilon _{\gamma}$ denotes the polarization vector of photon, and $P_L^{\mu}\equiv\gamma ^{\mu}(1-\gamma _5)$. The idea of factorization is to absorb the infrared (IR) behaviour into the wave-function, the matrix element can be written as the convolution of wave-function and hard kernel~\cite{Collinsmethod,Sachrajda}
\begin{equation}
\begin{split}
&F=\int dk \Phi(k)T_{\rm hard}(k)\\
\end{split}
\label{eq.1.2}
\end{equation}

For simplicity, we denote
\begin{equation}
\begin{split}
&x=m_Q^2,\;\;y=2k_Q\cdot p_{\gamma},\;\;z=2p_{\gamma}\cdot k_{\bar{q}},\;\;w=2k_Q\cdot k_{\bar{q}}\\
\end{split}
\label{eq.1.3}
\end{equation}
Using the wave-function defined in coordinate space
\begin{equation}
\Phi_{\alpha\beta}(x, y)=<0|\bar{q}_{\alpha}(x)[x,y]Q_{\beta}(y)|\bar{q}^S(p_{\bar{q}}),Q^s(p_Q)>
\label{eq.1.4}
\end{equation}
where $[x,y]$ is the Wilson Line which can be written as~\cite{Wilsonline}
\begin{equation}
\begin{split}
&[x,y]=\exp \left[ig_s\int _y^x d^4z z_{\mu} A^{\mu}(z)\right]=\sum _{\substack{n}}\frac{(ig_s)^n}{n!}\prod _{\substack{i}}^n\int _y^x d^4z_i z_{i\mu} A^{\mu}(z_i)
\label{eq.1.5}
\end{split}
\end{equation}
it has been proved that~\cite{factorization}, the matrix element up to the order $O(\left.\alpha _s\Lambda _{\rm QCD}/m_Q\right)$ at one-loop can be factorized as
\begin{equation}
\begin{split}
&F^{\mu}(\mu)=\sum _n \int d^4k_Q\int d^4 k_{\bar{q}}\Phi (k_Q, k_{\bar{q}}) C_n(k_Q, k_{\bar{q}},\mu) T_n(k_Q, k_{\bar{q}})\\
\end{split}
\label{eq.1.6}
\end{equation}
with
\begin{equation}
\begin{split}
&T_1=-e_q\frac{\slashed \varepsilon _{\gamma }^*\slashed p_{\gamma }}{2p_{\gamma}\cdot k_{\bar{q}}}P_L^{\mu },\;\;\;T_2=e_q\frac{2 \varepsilon _{\gamma }^*\cdot k_q}{2p_{\gamma}\cdot k_{\bar{q}}}P_L^{\mu },\;\;\;T_3=e_QP_L^{\mu }\frac{\slashed p_{\gamma}\slashed \varepsilon _{\gamma }^*}{2k_Q\cdot p_{\gamma}}\\
&T_4=e_QP_L^{\mu }\frac{m_Q\slashed \varepsilon _{\gamma }^*}{2k_Q\cdot p_{\gamma}},\;\;\;
T_5=\frac{-e\slashed p_{\gamma }\slashed \varepsilon _{\gamma }^*}{2p_{\gamma}\cdot p_l}P_L^{\mu }+eP_L^{\mu}\left(\frac{\varepsilon \cdot p_P}{p_{\gamma }\cdot p_P}-\frac{\varepsilon \cdot p_l}{p_{\gamma }\cdot p_l}\right)
\end{split}
\label{eq.1.7}
\end{equation}
except for $C_1 T_1$, all the other products are contribution of order $O(\Lambda_{\rm QCD}\left/m_Q\right.)$, for clarity, we define $C_1=C_1^0+C_1^1$, with $C_1^m$ represents order $O((\Lambda _{\rm QCD}\left/m_Q\right.)^m)$ contribution, the coefficients are
\begin{equation}
\begin{split}
&C_1^0=\left\{1+\frac{\alpha _s(m_Q)C_F}{4\pi}\left(-2{\rm Li}_2\left(1-\frac{y}{x}\right)-2\log^2\frac{y}{x}-\frac{y}{x-y}\log\frac{y}{x}+2\log\frac{y}{x}-6-\frac{\pi^2}{12}\right)\right\}\\
&\times \exp \left(\frac{\alpha _s(m_Q) C_f}{4\pi}\left(-4\log^2\frac{\mu}{m_Q}+4\log\frac{y}{x}\log\frac{\mu}{m_Q}-6\log\frac{\mu}{m_Q}\right)\right)\\
&\times \left(1+\frac{\alpha _s(\mu) C_f}{4\pi}\left( \log^2 \frac{\mu ^2}{z}- 3-\frac{\pi^2}{4}\right)\right)
\end{split}
\label{eq.1.8}
\end{equation}
\begin{equation}
\begin{split}
&C_1^1=\frac{w }{(x-y)^2 y}\left(x^2\left(-4\log\frac{xz}{y^2}-2\right)+xy\left(5\log\frac{x}{y}+8\log \frac{z}{y}+5\right)+y^2\left(2\log\frac{y}{x}+4\log\frac{y}{z}-3\right)\right)\\
&+\frac{x^3z}{(x-y)^2 y^2}\left(-3y_1+8\log\frac{xz}{y^2}-4\right)+\frac{x^2z}{(x-y)^2 y}\left(7y_1+13\log\frac{y}{x}+19\log\frac{y}{z}+14\right)\\
&\left.+\frac{xz}{(x-y)^2}\left(-5y_1+4\log\frac{x}{y}+14\log\frac{z}{y}-17\right)+\frac{yz}{(x-y)^2}\left(y_1-3\log\frac{z}{y}+2\log\frac{x}{y}+7\right)\right.\\
&\left.-\frac{w}{2(x-y)}\log\frac{x}{y}+\left(\frac{2w}{y}-\frac{4xz}{y^2}\right)(\log\frac{xz}{y^2}-\frac{1}{2}y_1)+\left(\frac{2w}{y}-\frac{4xz}{y^2}\right)\left(  \log \frac{x}{\mu^2}\log \frac{y}{z}-2\log \frac{y}{z}-\log\frac{x}{\mu^2}+4\right)\right.\\
&-\left(2-\frac{\pi^2}{3}-\log ^2\frac{y}{z} +\log \frac{y}{z} \log \frac{x}{\mu^2}+\log \frac{x}{\mu^2} \right)\frac{2z}{y}\\
\end{split}
\label{eq.1.9}
\end{equation}
\begin{equation}
\begin{split}
&C_2=1+\frac{\alpha _s C_F}{4\pi}\left(-\log \frac{y}{\mu^2}+y_1-2\log \frac{xz}{y^2}+\log\frac{x}{y}+\frac{xz}{yw}(\log\frac{xz}{y^2}-\frac{1}{2}y_1)\right.\\
&\left.-\frac{4zx}{yw}\left( \log \frac{x}{\mu^2}\log \frac{y}{z}-2\log \frac{y}{z}-\log\frac{x}{\mu^2}+4\right)\right)\\
\end{split}
\label{eq.1.10c2}
\end{equation}
\begin{equation}
\begin{split}
&C_3=1+\frac{\alpha _s C_F}{4\pi}\left(\left(\log \frac{x}{\mu^2}-4-\frac{x}{y}\left(-{\rm Li}_2(1-\frac{y}{x})+\frac{\pi^2}{3}\right)-\frac{6x-y}{x-y}\log \frac{x}{y}\right)\right.\\
&\left.-1+\log\frac{x}{\mu^2}-\frac{x}{x-y}+\frac{y(2x-y)}{(x-y)^2}\log\frac{x}{y}+\frac{2x}{y}(-5+3\log\frac{x}{\mu^2}+\frac{x}{x-y}+\frac{(2x-3y)y}{(x-y)^2}\log\frac{x}{y}) \right.\\
&\left.-\log\frac{y}{\mu^2}-2+\frac{x}{y-x}+\frac{x^2}{(x-y)^2}\log\frac{x}{y}+\frac{8w}{w-z}\log \frac{y}{w-z}\right)\\
\end{split}
\label{eq.1.10c3}
\end{equation}
\begin{equation}
\begin{split}
&C_4=\frac{\alpha _s C_F}{4\pi}\left(2\left(\frac{2xz^2}{yw}-2z\right)\frac{4w}{w-z}\log \frac{w-z}{y}\right.\\
&\left.-\frac{2y}{x-y}\log\frac{x}{y}+\left(-5+3\log\frac{x}{\mu^2}+\frac{x}{x-y}+\frac{y(2x-3y)}{(x-y)^2}\log\frac{x}{y}\right)\right)\\
\end{split}
\label{eq.1.11}
\end{equation}
where $y_1$ is defined as:
\begin{equation}
\begin{split}
&y_1=-\pi^2+2{\rm Li}_2\left(1-\frac{x}{y}\right)-\log^2\frac{xz}{y^2}+\log^2\frac{x}{y}\\
\end{split}
\label{eq.1.12}
\end{equation}
In the results given above, the large logarithms at order $O\left((\Lambda _{\rm QCD}/m_Q)^0\right)$ is already resummed. The $C_5 T_5$ term is the contribution from $\mathcal{A}_c$, and it dose not have 1-loop hard scattering kernel, so we have
\begin{equation}
\begin{split}
&C_5=1
\end{split}
\label{eq.1.13}
\end{equation}

The numerical result can be obtained by using the wave-function with the one obtained in Ref.~\cite{wavefunction}
\begin{equation}
\begin{split}
&\Phi (k_q, k_Q)=\frac{1}{\sqrt{3}}\int d^3 k \Psi (k) \frac{1}{\sqrt{2}}M\mid 0> \\
&\times \delta ^3(\vec{k}_{\bar{q}} + \vec{k})\delta ^3(\vec{k}_Q - \vec{k})\delta (k_{\bar{q}0}-\sqrt{k^2+m_q^2})\delta (k_{Q0}-\sqrt{k^2+m_Q^2})
\end{split}
\label{eq.1.14}
\end{equation}
with
\begin{equation}
\begin{split}
&M=\sum _{\substack{i}}b_Q^{i+}(\vec{k},\uparrow)d_q^{i+}(-\vec{k},\downarrow)-b_Q^{i+}(\vec{k},\downarrow)d_q^{i+}(-\vec{k},\uparrow), \;\;\Psi (\vec{k})=4\pi \sqrt{m_P \lambda _P ^3} e^{-\lambda _P |\vec{k}|}
\end{split}
\label{eq.1.15}
\end{equation}
and the result is
\begin{equation}
\begin{split}
&<\gamma\mid \bar{q} \Gamma ^{\mu} Q\mid P>=\epsilon _{\mu\nu\rho\sigma}\varepsilon ^{\nu}p_P^{\rho}p_{\gamma}^{\sigma}(F_V+F_{c1})\\
&+i\left(\varepsilon ^{\mu}p_P\cdot p_{\gamma}-p_{\gamma}^{\mu}\varepsilon \cdot p_P\right)(F_A+F_{c1})+F_{c2} p_P^{\mu}\\
\end{split}
\label{eq.1.16}
\end{equation}
with
\begin{equation}
\begin{split}
&F_V=\frac{1}{(2\pi)^3}\frac{3}{\sqrt{6}}\int d^3 k \Psi (k)\frac{1}{2\sqrt{p_{q0}p_{Q0}(p_{q0}+m_q)(p_{Q0}+m_Q)}} \frac{1}{m_PE_{\gamma}}\\
&\times \left(2e_qC_1m_Q-C_1^0p_{q0}e_q+e_Q 2p_{q0} C_3\right)\\
&F_A=\frac{1}{(2\pi)^3}\frac{3}{\sqrt{6}}\int d^3 k \Psi (k)\frac{1}{2\sqrt{p_{q0}p_{Q0}(p_{q0}+m_q)(p_{Q0}+m_Q)}} \frac{1}{m_PE_{\gamma}}\\
&\times \left(2e_qC_1m_Q-C_1^0p_{q0}e_q-e_q\frac{2p_{q0}m_Q}{E_{\gamma}}C_2-e_Q 2p_{q0} C_3+e_Q\frac{2p_{q0}m_Q}{E_{\gamma}}C_4\right)\\
\end{split}
\label{eq.1.17}
\end{equation}
where $F_{c1}$ and $F_{c2}$ come from the contribution of $\mathcal{A}_c$. Using
\begin{equation}
\begin{split}
&f_P p_P^{\mu}=i<0|\bar{q} \gamma ^{\mu} \gamma _5 Q|P>=-i<0|\bar{q} \gamma ^{\mu}(1- \gamma _5) Q|P>
\end{split}
\label{eq.1.18}
\end{equation}
we find
\begin{equation}
\begin{split}
&F_{c1}= \frac{-ef_P}{2p_{\gamma }\cdot p_l},\;\;\;F_{c2}= ief_Pp_P^{\mu}\left(\frac{\varepsilon \cdot p_P}{p_{\gamma }\cdot p_P}-\frac{\varepsilon \cdot p_l}{p_{\gamma }\cdot p_l}\right)
\end{split}
\label{eq.1.19}
\end{equation}
and the decay amplitude can be written as
\begin{equation}
\begin{split}
&\mathcal{A}_{SM}=\frac{eG_fV_{Qq}}{\sqrt{2}}<\gamma\mid \bar{q} \Gamma ^{\mu} Q\mid P>(\bar{l}P_L^{\mu}\nu)
\end{split}
\label{eq.1.20}
\end{equation}

\section{\label{sec:level3}2HDM contribution}

\begin{figure}
\includegraphics[scale=1.8]{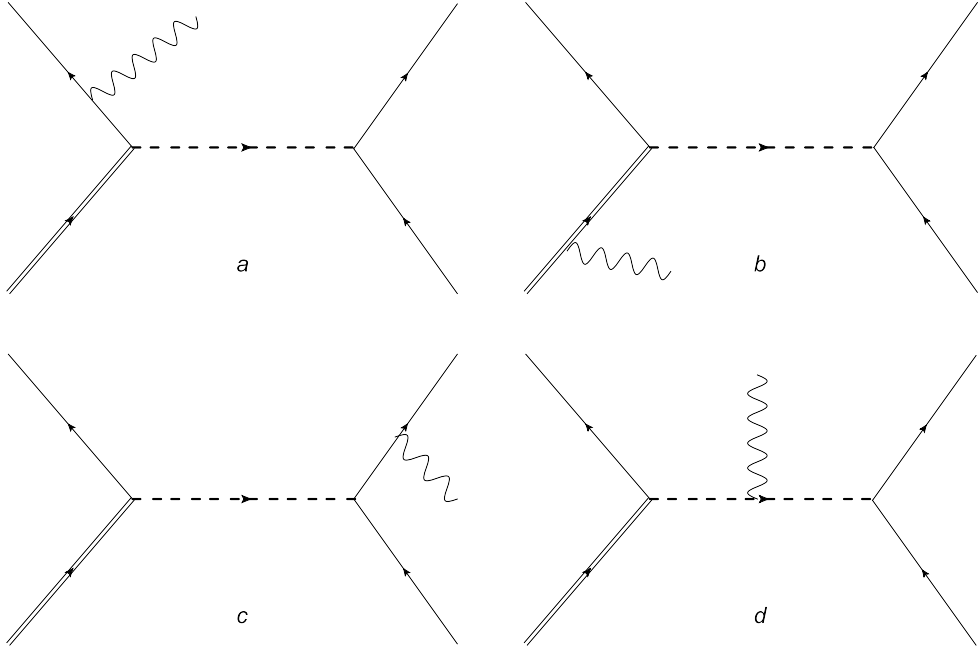}
\caption{\label{Tree2} Tree level amplitudes of charged Higgs bosons. The double line represents the heavy quark, and the dashed line, the charged Higgs.}
\end{figure}

The Feynman diagrams at tree level of the radiative leptonic decay in 2HDM are shown in Fig.~\ref{Tree2}. The contribution of Fig.~\ref{Tree2}.~d is suppressed by Higgs propagator and neglected, the decay amplitudes of the others can be written as
\begin{equation}
\begin{split}
&\mathcal{A}_a^{\rm 2HDM}=\frac{ie_q g^2 V_{Qq}m_Qm_l\tan ^2 \beta}{8M_W^2M_{H^+}^2} \bar{q}(p_{\bar{q}}) \slashed \varepsilon _{\gamma}^* \frac{\slashed p_{\gamma} - \slashed p_{\bar{q}}}{(p_{\gamma}-p_{\bar{q}})^2 -m_q^2}(1+\gamma _5)Q(p_Q) \left(\bar{l} (1-\gamma _5)\nu\right)\\
&\mathcal{A}_b^{\rm 2HDM}=\frac{ie_Q g^2 V_{Qq}m_Qm_l\tan ^2 \beta}{8M_W^2M_{H^+}^2} \bar{q}(p_{\bar{q}}) (1+\gamma _5)\frac{\slashed p_Q -\slashed p_{\gamma} +m_Q}{2p_Q\cdot p_{\gamma}} \slashed \varepsilon _{\gamma}^* Q(p_Q) \left(\bar{l} (1-\gamma _5)\nu\right)\\
&\mathcal{A}_c^{\rm 2HDM}=\frac{-i e g^2 V_{Qq}m_Qm_l\tan ^2 \beta}{8M_W^2M_{H^+}^2} \bar{q}(p_{\bar{q}})(1+\gamma _5)Q(p_Q) \left(\bar{l} \slashed \varepsilon _{\gamma}^* \frac{\slashed p_{\gamma} - \slashed p_l + m_l}{2p_l\cdot p_{\gamma}}(1-\gamma _5)\nu\right)\\
\end{split}
\label{eq.2.1}
\end{equation}
where $M_{H^+}$ is the mass of the charged Higgs boson, and $\tan \beta$ is a free parameter in 2HDM. The leading contribution at order $O((\left. \Lambda _{\rm QCD}/m_Q\right)^0)$ is $\mathcal{A}_a^{\rm 2HDM}$, which will vanish. This can be shown immediately, the matrix element can be written as
\begin{equation}
\begin{split}
&<\gamma |\bar{\nu}_{\bar{q}}\slashed \varepsilon _{\gamma}^* \frac{\slashed p_{\gamma} - \slashed p_{\bar{q}}}{(p_{\gamma}-p_{\bar{q}})^2 -m_q^2}(1+\gamma _5)u_Q|P> =<\gamma|\bar{\nu}_{\bar{q}}\slashed \varepsilon _{\gamma}^* \frac{\slashed p_{\gamma} - \slashed p_{\bar{q}}}{(p_{\gamma}-p_{\bar{q}})^2 -m_q^2}\gamma ^{\mu}(1-\gamma _5)\frac{ p_{Q\mu}}{m_Q}u_Q|P>\\
\end{split}
\label{eq.2.2}
\end{equation}
at the leading order, we can replace $\left. p_{Q\mu} / m_Q\right.$ by $\left. p_{P\mu} / m_P\right.$, and the matrix element can be factorized as~\cite{bsw1,Sachrajda}
\begin{equation}
\begin{split}
&<\gamma |\bar{\nu}_{\bar{q}} \slashed \varepsilon _{\gamma}^* \frac{\slashed p_{\gamma} - \slashed p_{\bar{q}}}{(p_{\gamma}-p_{\bar{q}})^2 -m_q^2}\gamma ^{\mu}(1-\gamma _5)u_Q|P>= i F_A \epsilon ^{\mu\nu\rho\sigma}\varepsilon _{\gamma \nu}p_{\gamma \rho}p_{P \sigma} + F_c(\varepsilon _{\gamma}^{\mu}p_P\cdot p_{\gamma} - p_{\gamma}^{\mu}\varepsilon _{\gamma}\cdot p_P)
\end{split}
\label{eq.2.3}
\end{equation}
Contracting with $p_{P\mu}$, it will vanish.

In fact, both Fig.~\ref{Tree2}.~a and Fig.~\ref{Tree2}.~b, and the QCD corrections of them will not contribute to the decay amplitude because of the symmetry of the wave-function of the pseudoscalar mason. Using the similar factorization procedure as in the SM, we find that, up to the order $O(\left.\alpha _s \Lambda _{\rm QCD}/m_Q\right)$, the matrix element of Fig.~\ref{Tree2}.~a and Fig.~\ref{Tree2}.~b with 1-loop QCD corrections can be factorized as
\begin{equation}
\begin{split}
&F_{\rm 2HDM}(\mu)=\sum _n \int d^4k_Q\int d^4 k_{\bar{q}}\Phi (k_Q, k_{\bar{q}}) C_n^{\rm 2HDM}(k_Q, k_{\bar{q}},\mu) T_n^{\rm 2HDM}(k_Q, k_{\bar{q}})\\
\end{split}
\label{eq.2.4}
\end{equation}
the delta-function in the wave-function will replace the transmission momenta $k_q$, $k_Q$ to the on-shell momenta $p_q$, $p_Q$. Using the definition of the wave-function in Eqs.~(\ref{eq.1.14}) and (\ref{eq.1.15}), in Dirac representation, we find
\begin{equation}
\begin{split}
&F_{\rm 2HDM}(\mu)=\sum _n \int d^3k \Psi (k) C_n^{\rm 2HDM}(x,\hat{y},\hat{z},\hat{w},\mu) Tr[M\cdot T_n^{\rm 2HDM}(p_Q, p_{\bar{q}})]\\
\end{split}
\label{eq.2.5}
\end{equation}
with
\begin{equation}
\begin{split}
&\hat{y}=2p_Q\cdot p_{\gamma},\;\;\;\hat{z}=2p_{\bar{q}}\cdot p_{\gamma},\;\;\;,\hat{y}=2p_Q\cdot p_{\bar{q}}\\
&M=\frac{(-\frac{1}{2})(\slashed p_Q + m_Q)(1+\gamma _0)\slashed p_{\bar{q}} \gamma _5}{\sqrt{p_{\bar{q}0}(p_{\bar{q}0}+m_q)p_{Q0}(p_{Q0}+m_Q)}}\\
\end{split}
\label{eq.2.6}
\end{equation}
and
\begin{equation}
\begin{split}
&T_1^{\rm 2HDM}=e_q\frac{\slashed \varepsilon _{\gamma} \slashed p_{\gamma}}{2p_{\bar{q}}\cdot p_{\gamma}}(1+\gamma _5),\;\;
T_2^{\rm 2HDM}=e_q\frac{\slashed \varepsilon _{\gamma} \slashed p_{\bar{q}}}{2p_{\bar{q}}\cdot p_{\gamma}}(1+\gamma _5),\;\;
T_3^{\rm 2HDM}=e_q\frac{\slashed \varepsilon _{\gamma} \slashed p_Q}{2p_{\bar{q}}\cdot p_{\gamma}}(1+\gamma _5)\\
&T_4^{\rm 2HDM}=e_Q(1+\gamma _5)\frac{\slashed p_{\gamma}}{2p_Q\cdot p_{\gamma}}\slashed \varepsilon _{\gamma} ,\;\;
T_5^{\rm 2HDM}=e_Q(1+\gamma _5)\frac{m_Q\slashed \varepsilon _{\gamma} }{2p_Q\cdot p_{\gamma}}\\
\end{split}
\label{eq.2.7}
\end{equation}
The kinematic indicates that
\begin{equation}
\begin{split}
&p_{\gamma}=(E_{\gamma},0,0,-E_{\gamma}),\;\;\; \varepsilon = (0, \varepsilon _1, \varepsilon _2, 0)\\
&p_{\bar{q}}=(p_{\bar{q}0}, k_1, k_2, k_3),\;\;\; p_Q=(p_{Q0}, -k_1, -k_2, -k_3)
\end{split}
\label{eq.2.8}
\end{equation}
Notice that, $C_n^{\rm 2HDM}$ are either even functions of $k_1$ or $k_2$, in this case, after the trace, all terms are vanished, or the odd function of $k_1$ or $k_2$, then, all terms will vanish after the integral $\int d^3 k$.

However, $\mathcal{A}_c^{\rm 2HDM}$ will survive and will contribute to the decay amplitude. As the same case in the SM, $\mathcal{A}_c^{\rm 2HDM}$ also dose not receive contribution from 1-loop hard scattering kernel. As a result, we find
\begin{equation}
\begin{split}
&\mathcal{A}_{2HDM}=\frac{e G_f}{\sqrt{2}}\frac{\tan ^2 \beta m_l m_Q}{M_{H^+}^2}V_{Qq}<0|\bar{q}\Gamma Q|P>\bar{l}\slashed \varepsilon _{\gamma}^* \frac{\slashed p_{\gamma} - \slashed p_l + m_l}{2p_l\cdot p_{\gamma}}(1-\gamma _5)\nu\\
&=\frac{e G_f}{\sqrt{2}}\frac{\tan ^2 \beta m_l m_Q}{M_{H^+}^2}V_{Qq} f_{\rm 2HDM}\bar{l}\slashed \varepsilon _{\gamma}^* \frac{\slashed p_{\gamma} - \slashed p_l + m_l}{2p_l\cdot p_{\gamma}}(1-\gamma _5)\nu\\
\end{split}
\label{eq.2.9}
\end{equation}
with
\begin{equation}
\begin{split}
&f_{\rm 2HDM}=\frac{1}{(2\pi)^3}\frac{3}{\sqrt{6}}\int d^3 k \Psi (k)\frac{1}{2\sqrt{p_{q0}p_{Q0}(p_{q0}+m_q)(p_{Q0}+m_Q)}} \left(-2m_Q p_{\bar{q}0}-2p_{\bar{q}}\cdot p_Q\right)\\
\end{split}
\label{eq.2.10}
\end{equation}

\section{\label{sec:level4}Numerical result and analysis of parameters of 2HDM}

The form factors defined in Eq.~(\ref{eq.1.16}) can be calculated using the integral defined in Eq.~(\ref{eq.1.17}), and we evaluate this integral using \cite{wavefunction}
\begin{equation}
\begin{split}
&m_D=1.9\;{\rm GeV},\;m_B=5.1\;{\rm GeV},\;m_u=m_d=0.08\;{\rm GeV}\\
&m_b=4.98\;{\rm GeV},\;m_c=1.54\;{\rm GeV}\\
&\Lambda _{\rm QCD}=200\;{\rm MeV},\;\lambda _B = 2.8\;{\rm GeV}^{-1},\;\lambda _D = 3.4\;{\rm GeV}^{-1}
\end{split}
\label{eq.3.1}
\end{equation}

The numerical results of the form factors are inconvenient to use when calculate the decay widths. For simplicity, we use some simple forms to fit the numerical results. Inspired by the form factors in Ref.~\cite{naivefactorization}, the form factors are fitted as
\begin{equation}
\begin{split}
&F_{A,V}(E_{\gamma})=\left(A_{A,V}\frac{\Lambda_{\rm QCD}}{E_{\gamma}}+B_{A,V}\left(\frac{\Lambda_{\rm QCD}}{{E_{\gamma}}}\right)^2\right)
\end{split}
\label{eq.3.2}
\end{equation}
The results are more reliable at the region $E_{\gamma}\gg \Lambda _{\rm QCD}$ because we have neglected the higher order terms of $\Lambda _{\rm QCD}\left/E_{\gamma}\right.$. As we have done in Ref.~\cite{factorization}, we choose the region $E_{\gamma}>2\Lambda_{\rm QCD}$ to fit the parameters in Eq.~(\ref{eq.3.2}). The result of $B$ mason in the SM is
\begin{equation}
\begin{split}
&F_A^B(E_{\gamma})=\left(0.25\frac{\Lambda_{\rm QCD}}{E_{\gamma}}+0.39\left(\frac{\Lambda_{\rm QCD}}{{E_{\gamma}}}\right)^2\right)\;{\rm GeV}^{-1}\\
&F_V^B(E_{\gamma})=\left(0.28\frac{\Lambda_{\rm QCD}}{E_{\gamma}}-0.73\left(\frac{\Lambda_{\rm QCD}}{{E_{\gamma}}}\right)^2\right)\;{\rm GeV}^{-1}\\
\end{split}
\label{eq.3.3}
\end{equation}
For $D$ mason, there is an additional minus sign in $F_A$ and $F_V$, and the result is
\begin{equation}
\begin{split}
&F_A^D(E_{\gamma})=\left(-0.10\frac{\Lambda_{\rm QCD}}{E_{\gamma}}+0.76\left(\frac{\Lambda_{\rm QCD}}{{E_{\gamma}}}\right)^2\right)\;{\rm GeV}^{-1}\\
&F_V^D(E_{\gamma})=\left(0.39\frac{\Lambda_{\rm QCD}}{E_{\gamma}}+0.04\left(\frac{\Lambda_{\rm QCD}}{{E_{\gamma}}}\right)^2\right)\;{\rm GeV}^{-1}\\
\end{split}
\label{eq.3.4}
\end{equation}

\begin{figure}
\includegraphics[scale=0.8]{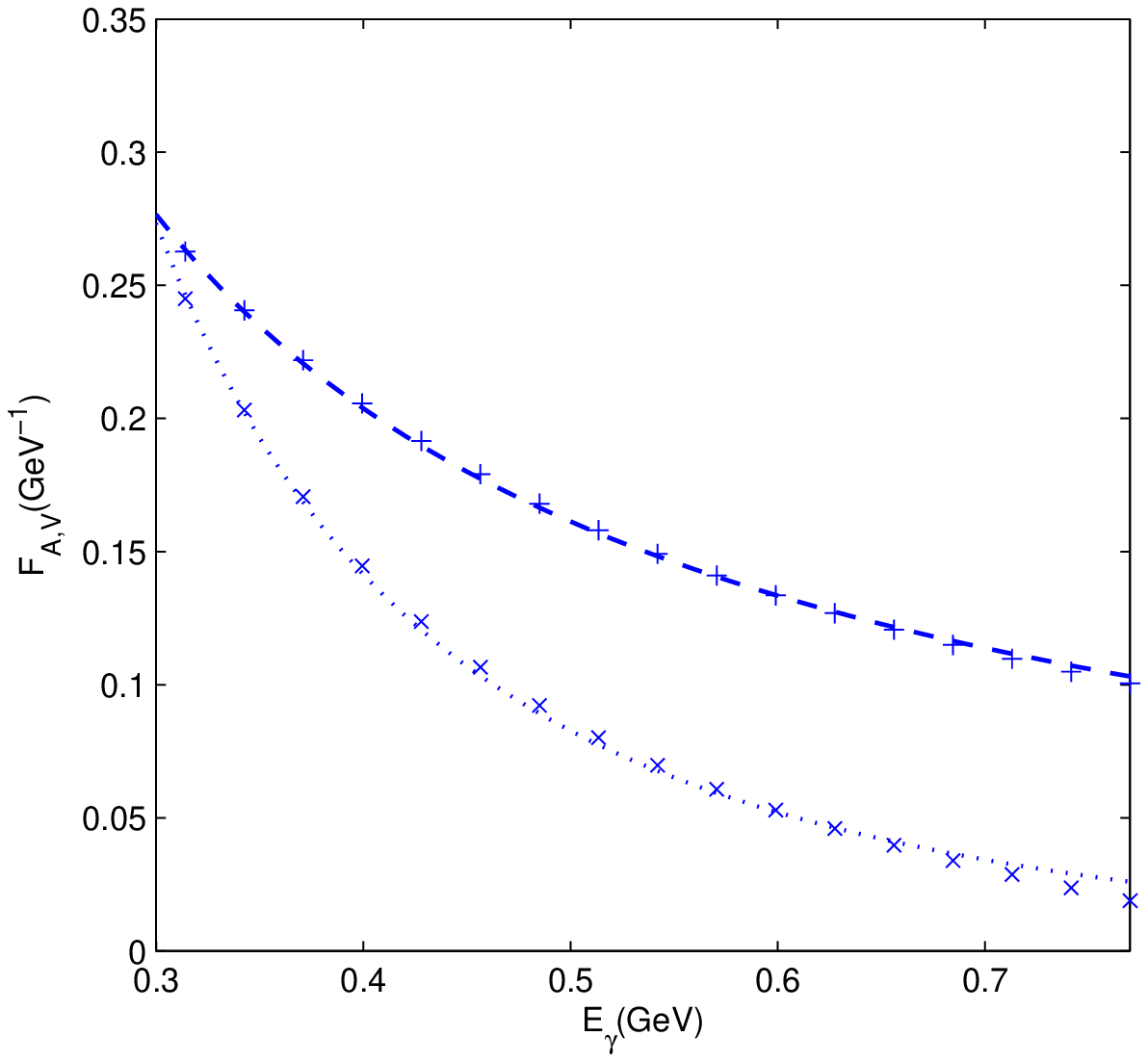}
\caption{\label{formb} Fit of the form factors of $B\to \gamma l \nu _l$. The points `$\times$' and `+' are the numerical results for the form factors $F_V$ and $F_A$, and the dotted and dashed curves are the fitted results using Eq.~(\ref{eq.3.2}).}
\end{figure}

\begin{figure}
\includegraphics[scale=0.8]{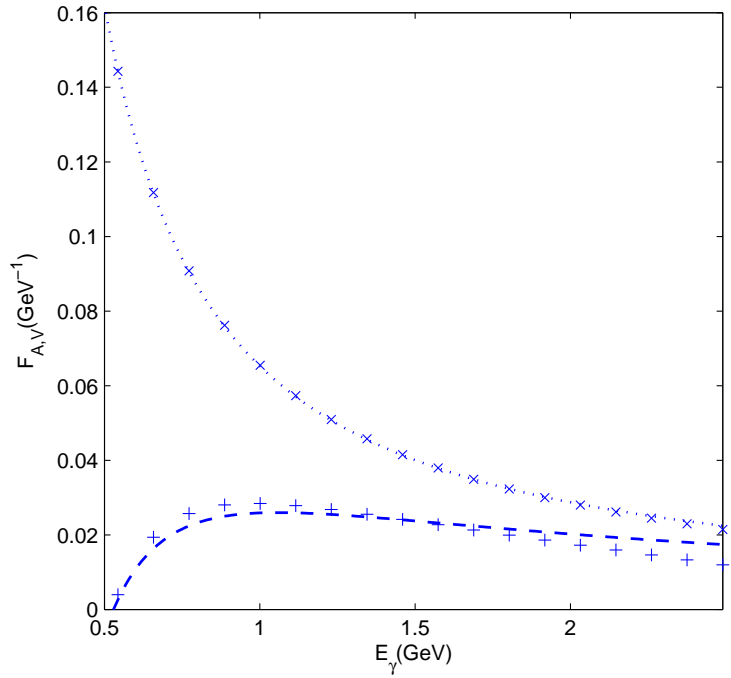}
\caption{\label{formd} Fit of the form factors of $D\to \gamma l \nu _l$. The points `$\times$' and `+' are the numerical results for the form factors $F_V$ and $F_A$, and the dotted and dashed curves are the fitted results using Eq.~(\ref{eq.3.2}).}
\end{figure}

The fitting of the form factors are shown in Fig.~\ref{formb} and Fig.~\ref{formd}.

On the other hand, $F_{c1}$ and $F_{c2}$ are related to the decay constant, and we use the result in Refs.~\cite{wavefunction,naivefactorization} which is calculated using the same wave-function and the same parameters as what we use in this work. The decay constants are
\begin{equation}
\begin{split}
&f_B=193.57\;{\rm MeV},\;\;\;f_D=204.98\;{\rm MeV}\\
\end{split}
\label{eq.3.5}
\end{equation}

The contribution of the 2HDM $f_{\rm 2HDM}$ can be calculated using Eq.~(\ref{eq.2.10}), and the result is
\begin{equation}
\begin{split}
&f_{\rm 2HDM}^B=-1.23\;{\rm GeV}^2,\;\;\;f_{\rm 2HDM}^D=-0.51\;{\rm GeV}^2
\end{split}
\label{eq.3.6}
\end{equation}

With $R$ defined as $R=\tan \beta \left./ M_{H\pm}\right.$, the branch ratios can be written as
\begin{equation}
\begin{split}
&Br _{\rm tot}= Br _{\rm SM}\times (1+R^2 m_Q m_l a  + R^4 m_Q^2 m_l^2 b )\\
\end{split}
\label{eq.3.7}
\end{equation}
with
\begin{equation}
\begin{split}
&a =\frac{1}{R^2 m_Q m_l}\frac{\mathcal {A}_{\rm SM}^2}{\mathcal {A}_{\rm SM}^*\mathcal {A}_{\rm 2HDM}+\mathcal {A}_{\rm 2HDM}^*\mathcal {A}_{\rm SM}},\;\;\;
b =\frac{1}{R^4 m_Q^2 m_l^2}\frac{\mathcal {A}_{\rm SM}^2}{\mathcal {A}_{\rm 2HDM}^2}\\
\end{split}
\label{eq.3.8}
\end{equation}

Just as the case of the pure leptonic decay~\cite{RConstraint1}, the interference term in the radiative leptonic decay is also found to be destructive. We calculate $Br_{\rm SM}$, $a$ and $b$ separately. Using the fitted result of $F_A$, $F_V$, the result of $f_P$ and $f_{\rm 2HDM}$, and using the Cabibbo-Kobayashi-Maskawa (CKM) matrix elements~\cite{particaldatagroup,vub}, and mass of the leptons
\begin{equation}
\begin{split}
&V_{\rm cd}=0.226,\;\;V_{\rm ub}=0.0047,\;\;m_{\tau}=1776.82\;{\rm MeV},\;\;m_{\mu}=105.658\;{\rm MeV}\\
\end{split}
\label{eq.3.9}
\end{equation}
we can obtain $Br _{\rm SM}$, $a $ and $b $. There are IR divergences in the radiative leptonic decays when the photon is soft or collinear with the emitted lepton. Theoretically this IR divergences can be canceled by adding the decay rate of the radiative leptonic decay with the pure leptonic decay rate, in which one-loop correction is included \cite{changch}. This is because the radiative leptonic decay can not be distinguished from the pure leptonic decay in experiment when the photon energy is smaller than the experimental resolution to the photon energy. So the decay rate of the radiative leptonic decay depend on the experimental resolution to the photon energy $E_{\gamma}$ which is denoted by $\Delta E_{\gamma}$. The dependence of the branching ratios of $B$ meson in the SM on the resolution are listed in Table~\ref{bsm}. And for the same reason, $a$ and $b$ also depend on $\Delta E_{\gamma}$, the results of $a$ and $b$ of B meson are listed in Table~\ref{Bab}.

\begin{table}[ht]
\begin{tabular}{ c c c| c c c }
\hline
    $\Delta E_{\gamma}$ & $Br_{\rm SM}(B\to \mu\nu _{\mu} \gamma)$ & $Br_{\rm SM}(B\to \tau\nu _{\tau} \gamma)$ & $\Delta E_{\gamma}$ & $Br_{\rm SM}(B\to \mu \nu _{\mu} \gamma)$ & $Br_{\rm SM}(B\to \tau\nu _{\tau} \gamma)$\\
\hline
  $ 5{\rm MeV}$ & $1.35\times 10^{-6}$ & $1.69\times 10^{-6}$ & $20{\rm MeV}$ & $1.13\times 10^{-6}$ & $1.27\times 10^{-6}$\\
  $10{\rm MeV}$ & $1.24\times 10^{-6}$ & $1.48\times 10^{-6}$ & $25{\rm MeV}$ & $1.10\times 10^{-6}$ & $1.20\times 10^{-6}$\\
  $15{\rm MeV}$ & $1.18\times 10^{-6}$ & $1.35\times 10^{-6}$ & $30{\rm MeV}$ & $1.07\times 10^{-6}$ & $1.14\times 10^{-6}$\\
  \hline
\end{tabular}
\caption{The branching ratios with different photon resolution $\Delta E_{\gamma}$ in the SM.}
\label{bsm}
\end{table}

\begin{table}[ht]
\begin{tabular}{ c |c c| c c| c |c c| c c }
\hline
    $\Delta E_{\gamma}$ & $a ^{B\to \mu\nu _{\mu} \gamma}$ & $b ^{B\to \mu\nu _{\mu} \gamma}$ & $a ^{B\to \tau\nu _{\tau} \gamma}$ & $b ^{B\to \tau\nu _{\tau} \gamma}$ & $\Delta E_{\gamma}$ & $a ^{B\to \mu\nu _{\mu} \gamma}$ & $b ^{B\to \mu\nu _{\mu} \gamma}$ & $a ^{B\to \tau\nu _{\tau} \gamma}$ & $b ^{B\to \tau\nu _{\tau} \gamma}$\\
\hline
  $ 5{\rm MeV}$ & $-7.47$ & $14.11$ & $-2.53$ & $12.49$ & $20{\rm MeV}$ & $-8.50$ & $10.01$ & $-2.47$ & $10.59$\\
  $10{\rm MeV}$ & $-7.98$ & $12.25$ & $-2.51$ & $11.68$ & $25{\rm MeV}$ & $-8.65$ & $9.19$ & $-2.46$ & $10.15$\\
  $15{\rm MeV}$ & $-8.28$ & $10.99$ & $-2.49$ & $11.09$ & $30{\rm MeV}$ & $-8.77$ & $8.49$ & $-2.45$ & $9.75$\\
  \hline
\end{tabular}
\caption{$a$ and $b$ defined in Eq.~(\ref{eq.3.8}) with different photon resolution $\Delta E_{\gamma}$.}
\label{Bab}
\end{table}

The result of $D\to \tau \nu _{\tau}\gamma$ is too small because of the phase space suppression, so we only calculate the branching ratios of $D\to \mu \nu _{\mu}\gamma$, and the result is listed in Table~\ref{dres}.
\begin{table}[ht]
\begin{tabular}{ c |c| cc | c| c|cc }
\hline
    $\Delta E_{\gamma}$ & $Br_{\rm SM}(D\to \mu\nu _{\mu} \gamma)$ & $a^{D\to \mu\nu _{\mu} \gamma} $ & $b^{D\to \mu\nu _{\mu} \gamma} $ & $\Delta E_{\gamma}$ & $Br_{\rm SM}(D\to \mu \nu _{\mu} \gamma)$ & $a^{D\to \mu\nu _{\mu} \gamma} $ & $b^{D\to \mu\nu _{\mu} \gamma} $ \\
\hline
  $ 5{\rm MeV}$ & $4.42\times 10^{-5}$ & $-7.00$ & $31.84$ & $20{\rm MeV}$ & $2.87\times 10^{-5}$ & $-9.47$ & $25.69$\\
  $10{\rm MeV}$ & $3.64\times 10^{-5}$ & $-8.09$ & $29.46$ & $25{\rm MeV}$ & $2.62\times 10^{-5}$ & $-9.97$  & $23.98$\\
  $15{\rm MeV}$ & $3.18\times 10^{-5}$ & $-8.86$ & $27.49$ & $30{\rm MeV}$ & $2.42\times 10^{-5}$ & $-10.38$ & $22.31$\\
  \hline
\end{tabular}
\caption{The branching ratios with different photon resolution $\Delta E_{\gamma}$ of $D\to \gamma \mu \nu _{\mu}$ in the SM, and the contribution of 2HDM indicated by $a$ and $b$ defined in Eq.~(\ref{eq.3.8}).}
\label{dres}
\end{table}

Using $\Delta E_{\gamma} = 10{\rm MeV}$~\cite{photonres}, the $Br_{\rm tot}$ can be written as
\begin{equation}
\begin{split}
&Br_{ B\to \gamma \mu \nu _{\mu}}=1.24 \times 10^{-6} \times r_H=1.24 \times 10^{-6} \times (1-4.20 R^2+3.39 R^4)\\
&Br_{ B\to \gamma \tau \nu _{\tau}}=1.48 \times 10^{-6} \times r_H=1.48 \times 10^{-6} \times (1-22.20 R^2 + 914.66 R^4)\\
&Br_{ D\to \gamma \mu \nu _{\mu}}=3.64 \times 10^{-5} \times r_H=3.64 \times 10^{-5} \times (1-1.32 R^2+0.78 R^4)\\
\end{split}
\label{eq.3.10}
\end{equation}
with $r_H$ defined as $Br_{\rm tot}\left./Br_{\rm SM}\right.$. The numerical results rely on $R$. In Ref.~\cite{RConstraint1}, two allowed regions for $R$ is given as
\begin{equation}
\begin{split}
&0<R_1<0.15\;{\rm GeV}^{-1},\;\;\;0.22{\rm GeV}^{-1}<R_2<0.33\;{\rm GeV}^{-1}
\end{split}
\label{eq.3.11}
\end{equation}
In Ref.~\cite{mhConstraint1}, the constraint for $M_{H^+}$ is $M_{H^+}>360\;{\rm GeV}$ at $95\%$ CL, and Ref.~\cite{mhConstraint2} shows that $M_{H^+}>315\;{\rm GeV}$ at $95\%$ CL.. In Ref.~\cite{tanbetaConstraint}, the $\tan \beta$ is also constrained. In the case that $m_h=126\;{GeV}$, $\tan \beta < 5$, while in the case that $m_h<126\;{GeV}, M_H=126\;{\rm GeV}$, $\tan \beta < 30$. Considering those constraints, we find $R<0.1\;{\rm GeV}$, and the $R_2$ region can be excluded.

\begin{figure}
\includegraphics[scale=0.75]{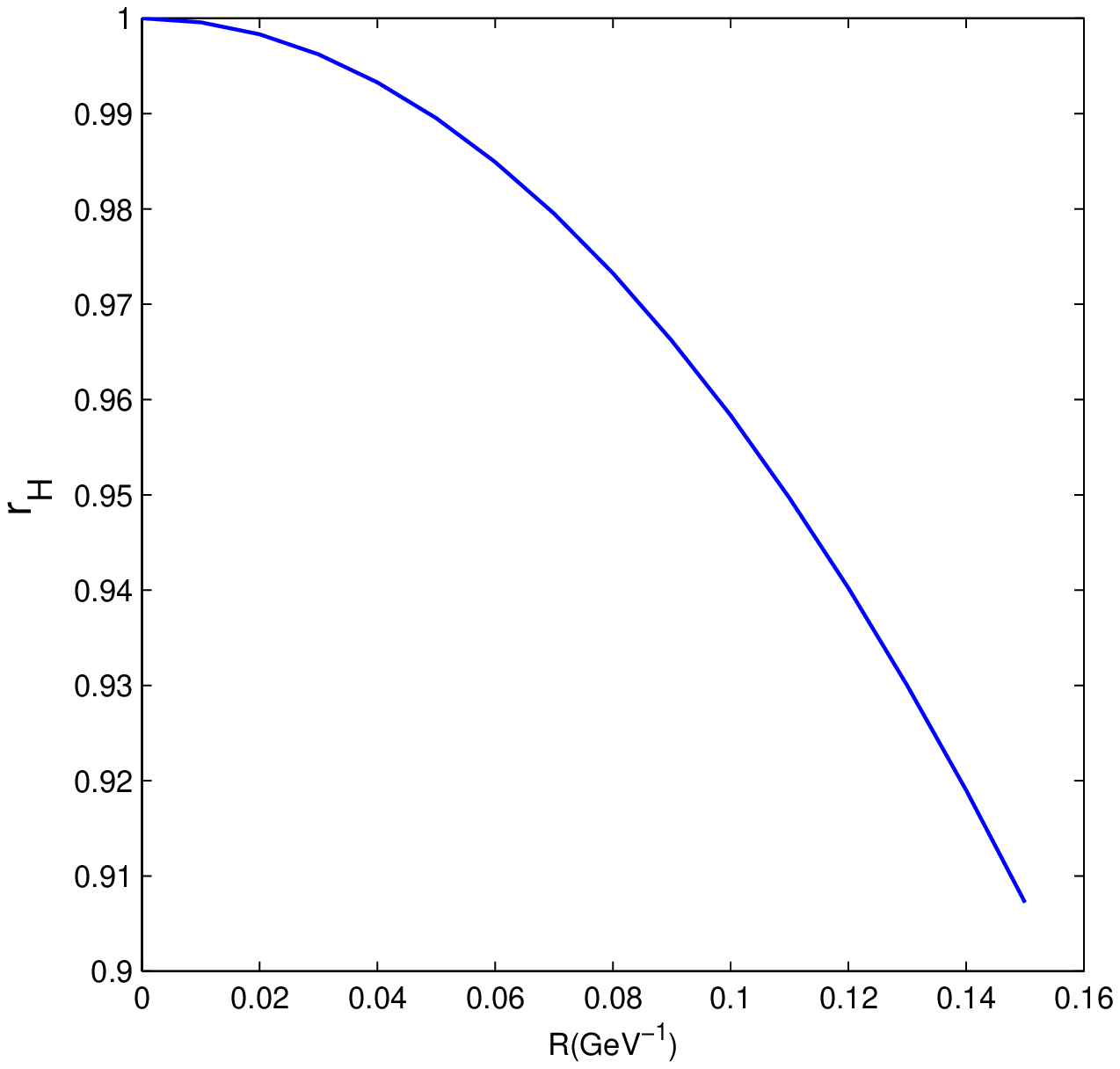}
\caption{\label{rhbmu} $r_H$ as function of $R$ in the $B\to \gamma \mu \nu _{\mu}$ decay mode.}
\end{figure}

\begin{figure}
\includegraphics[scale=0.75]{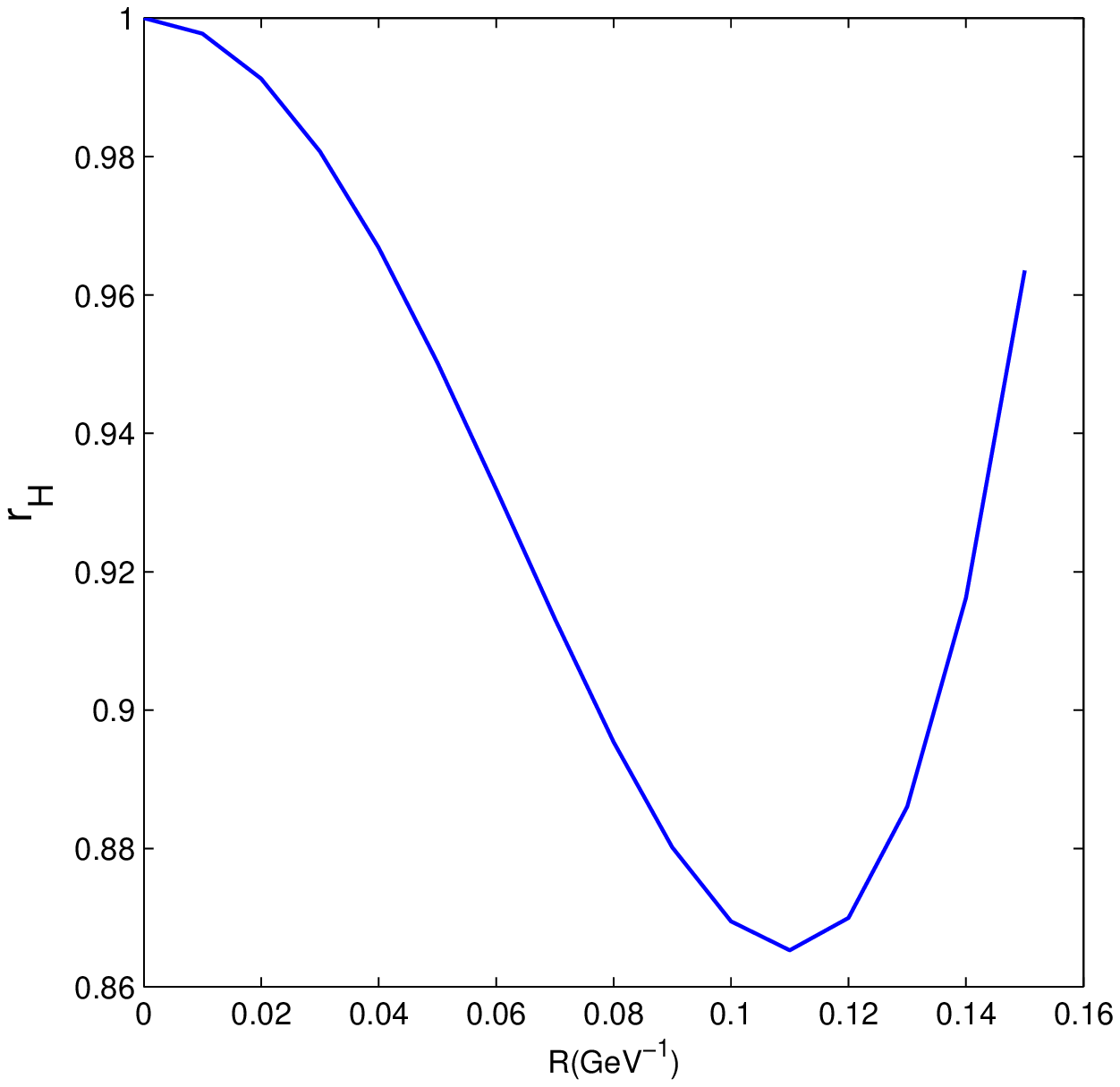}
\caption{\label{rhbtau} $r_H$ as function of $R$ in the $B\to \gamma \tau \nu _{\tau}$ decay mode.}
\end{figure}

\begin{figure}
\includegraphics[scale=0.75]{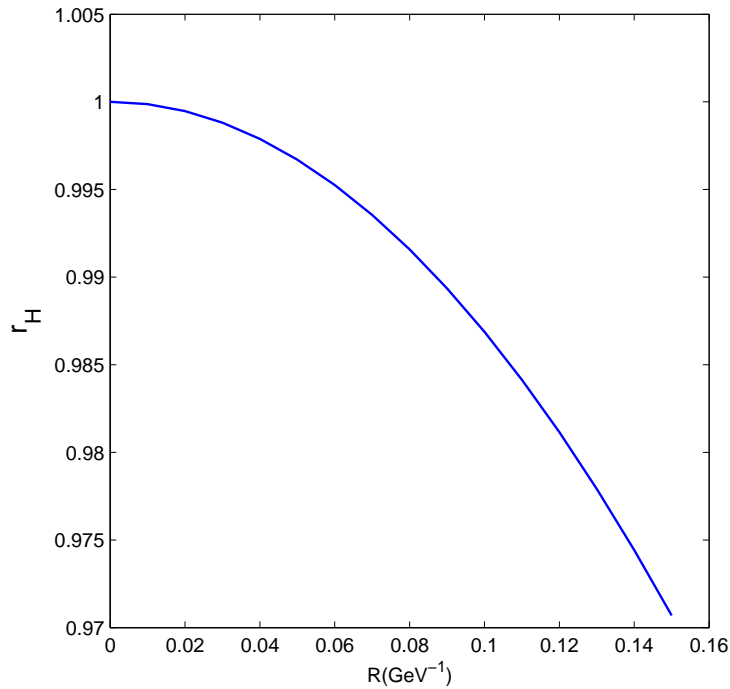}
\caption{\label{rhdmu} $r_H$ as function of $R$ in the $D\to \gamma \mu \nu _{\mu}$ decay mode.}
\end{figure}

The branching ratios including the contribution of charged Higgs bosons in the region that $0<R<0.15\;{\rm GeV}^{-1}$ are listed in Table~\ref{brtot}.
\begin{table}[ht]
\begin{tabular}{ c | c }
\hline
  &  $Br$ in $0<R<0.15\;{\rm GeV}^{-1}$ \\
\hline
 $B\to \gamma \mu \nu _{\mu}$ & $ 1.12\times 10^{-6}<Br_{\rm tot} < 1.24 \times 10^{-6}$ \\
 $B\to \gamma \tau \nu _{\tau}$ & $ 1.28\times 10^{-6}<Br_{\rm tot} < 1.48 \times 10^{-6}$\\
 $D\to \gamma \mu \nu _{\mu}$ & $ 3.53\times 10^{-5}<Br_{\rm tot} < 3.64\times 10^{-5} $\\
\hline
\end{tabular}
\caption{The branching ratios in the region $0<R<0.15\;{\rm GeV}^{-1}$}
\label{brtot}
\end{table}
And the ratio $r_H$ as a function of $R$ are shown in Figs.~\ref{rhbmu}, \ref{rhbtau} and \ref{rhdmu}.

We find that, the decay mode $B\to \gamma \tau \nu_{\tau}$ is very sensitive to the contribution of the charged Higgs bosons. For $R=0.1\;{\rm GeV^{-1}}$, for example, the branching ratios is suppressed by more than $13 \%$.

\section{\label{sec:level5}Summary}

In this paper, we calculated the branching ratios of the radiative leptonic decay of the heavy pseudoscalar meson with a massive lepton. The contribution of 2HDM-Type-II is included. The SM contribution is obtained using the factorization procedure up to the order $O(\left.\alpha _s \Lambda _{\rm QCD}/\right.m_Q)$ with one-loop correction. The contribution of the charged Higgs boson is also obtained by the factorization scheme, however, we find that only the diagram with the photon emitting from the lepton leg will contribute, which is an order $O(\left. \Lambda _{\rm QCD}/\right.m_Q)$ contribution and dose not receive contributions from 1-loop hard scattering kernel. The numerical results of the branching ratios are listed in Table~\ref{brtot}, and dependence of $r_H$ on $R$ is shown in Figs.~\ref{rhbmu}, \ref{rhbtau} and \ref{rhdmu}.

We find that, the decay mode $B\to \gamma \tau \nu_ {\tau}$ is sensitive to the contribution of the charged Higgs in the 2HDM. This decay mode is as sensitive as the pure leptonic decay of $B$ meson, which is estimated to be $r_H=(1-m_P^2 R^2)$~\cite{RConstraint1}. However, the result of the pure leptonic decay is derived at the leading order of $O((\left. \Lambda _{\rm QCD}/m_Q\right)^0)$, while our result is calculated up to the order $O(\left. \alpha _s \Lambda _{\rm QCD}/m_Q\right)$. On the other hand, the branching ratios of this decay mode in the SM is about $1.48\times 10^{-6}$, which is larger than $B \to e\nu _e$ and $B\to \mu\nu _{\mu}$ both are believed to be less than $10^{-6}$~\cite{particaldatagroup}. The decay mode $B\to \gamma \mu \nu _{\mu}$ is also important because the charged Higgs can suppress the branching ratios as large as about $10\%$. It is more sensitive than the pure leptonic decay modes $D\to l \nu$, which is estimated to be $r_H=(1-m_P^2 R^2)=(1-3.6 R^2)$. Though the branching ratio of the radiative leptonic decay of the $B$ meson is much smaller, the result in the SM is more reliable than the case of the $D$ meson because the heavy quark mass $m_b$ is larger then $m_c$.

We also find that, the decay mode $D\to \mu \nu _{\mu}\gamma$ is rarely affected by the charged Higgs boson. On the other hand, the numerical result of this mode is also not sufficiently accurate in the SM because the mass of $c$ quark is not so heavy. This decay mode should be excluded in the search of the charged Higgs bosons.


\textbf{Acknowledgements}: This work is supported in part by the
National Natural Science Foundation of China under contracts Nos.
11375088, 10975077, 10735080, 11125525.

\end{document}